\documentclass[10pt,a4paper]{scrartcl}

\usepackage[top=1.5cm, bottom=2.5cm, left=2.5cm, right=2.5cm]{geometry}
\usepackage{multicol}
\usepackage{float}
\usepackage{authblk}
\usepackage{graphicx}
\usepackage{subfig}
\usepackage{etoolbox} 
\robustify{\subref}
\usepackage[binary-units=true]{siunitx}
\usepackage{hyperref}

\setkomafont{publishers}{\small}

\begin{document}

\title{BrainScaleS Large Scale Spike Communication using Extoll}
\date{November 11, 2019}
\publishers{Extended Abstract, submitted to \textit{Neuro Inspired Computational Elements 2020 (NICE'2020)},\\ accepted and presented as a poster in March 2021}

\author[1]{Tobias Thommes}

\author[2]{Niels Buwen}

\author[1]{Andreas Grübl}

\author[1]{Eric Müller}

\author[2]{Ulrich Brüning}

\author[1]{Johannes Schemmel}

\affil[1]{Kirchhoff-Institute for Physics, University of Heidelberg, Germany}
\affil[2]{Institute of Computer Engineering, University of Heidelberg, Germany}

\maketitle

\begin{multicols}{2}
\section*{Abstract}

The BrainScaleS Neuromorphic Computing System is currently connected to a compute cluster via Gigabit-Ethernet network technology.
This is convenient for the currently used experiment mode, where neuronal networks cover at most one wafer module.
When modelling networks of larger size, as for example a full sized cortical microcircuit model,
one has to think about connecting neurons across wafer modules to larger networks.
This can be done, using the Extoll networking technology, which provides high bandwidth and low latencies,
as well as a low overhead packet protocol format.

\section{Introduction}

\begin{figure}[H]
    \centering
    \includegraphics[width=0.95\linewidth]{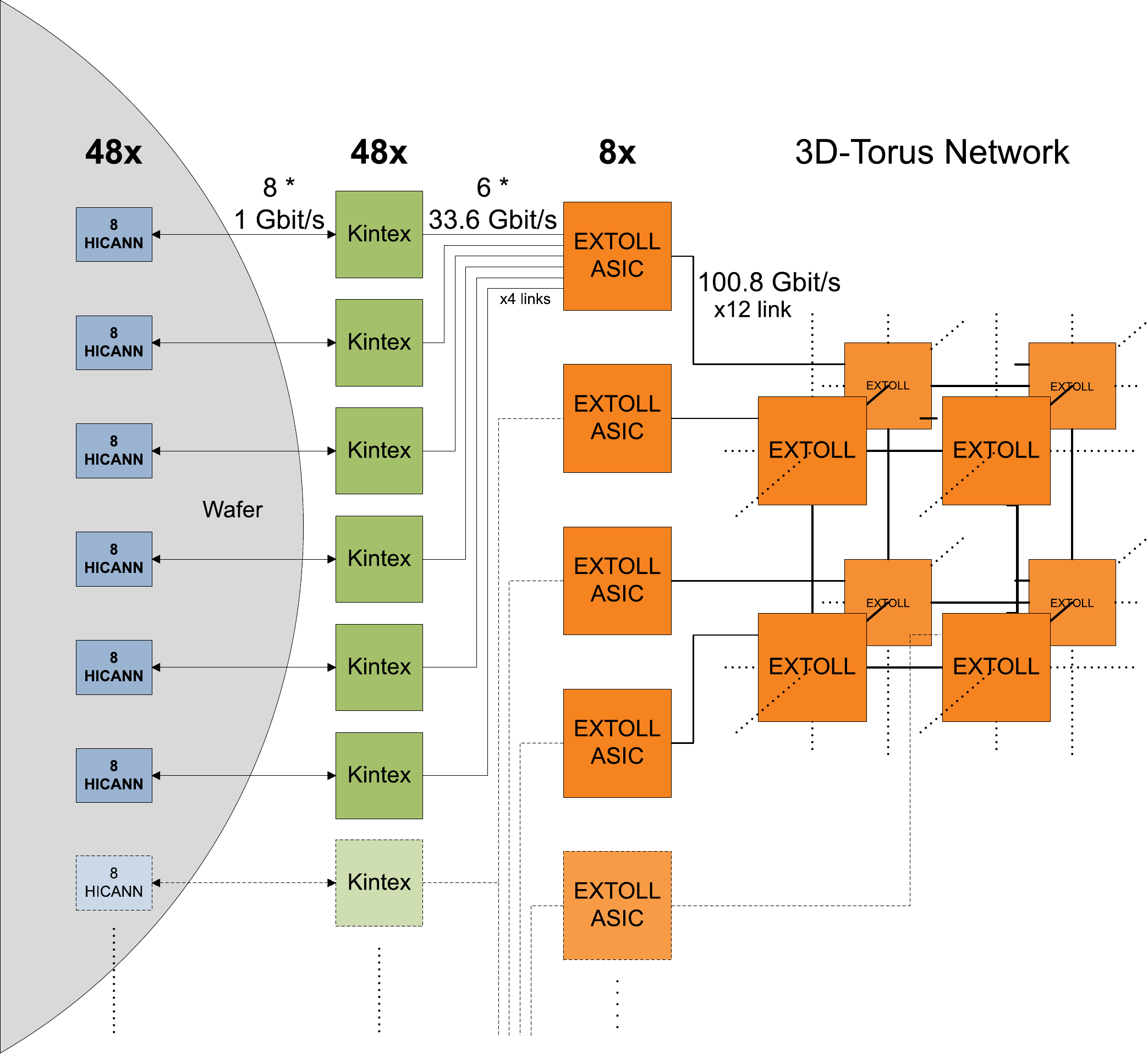}
    \caption{The BrainScaleS Extoll Network topology.}
    \label{fig:hbp_extoll_network}
\end{figure}

The Extoll network technology \cite{nussle2009fpga} is based on the Tourmalet Network Interface Card (NIC).
It offers \num{7} links and implements all the switching and interfacing capabilities, necessary to build an HPC network.
Each Extoll link can comprise up to \num{12} serial lanes of \SI{8.4}{\giga\bit\per\second} each.
The NIC can be connected to a host node through a PCIe~x16~Gen3 connector.
In an Extoll network, the nodes are usually connected in a 3D-Torus topology, which offers good scaling characteristics.
Routing of messages through the network is entirely done by the Tourmalet network chips and is based on a given \SI{16}{\bit} destination address in the message header.
A BrainScaleS wafer module \cite{schemmel2010wafer} contains \num{48} reticles,
each reticle comprising \num{8} HICANN chips which are connected to a Kintex~7 FPGA through \num{8} \SI{1}{\giga\bit\per\second} serial links.
We consider the topology shown in Figure~\ref{fig:hbp_extoll_network} to be optimal for interconnecting multiple wafers regarding bandwidth utilisation.
\num{6} of these FPGAs are gathered at one of \num{8} concentrator nodes per wafer module, connecting them to one torus node, respectively.

\begin{figure*}[t]
    \centering
    \subfloat[]{
        \includegraphics[width=0.3\textwidth]{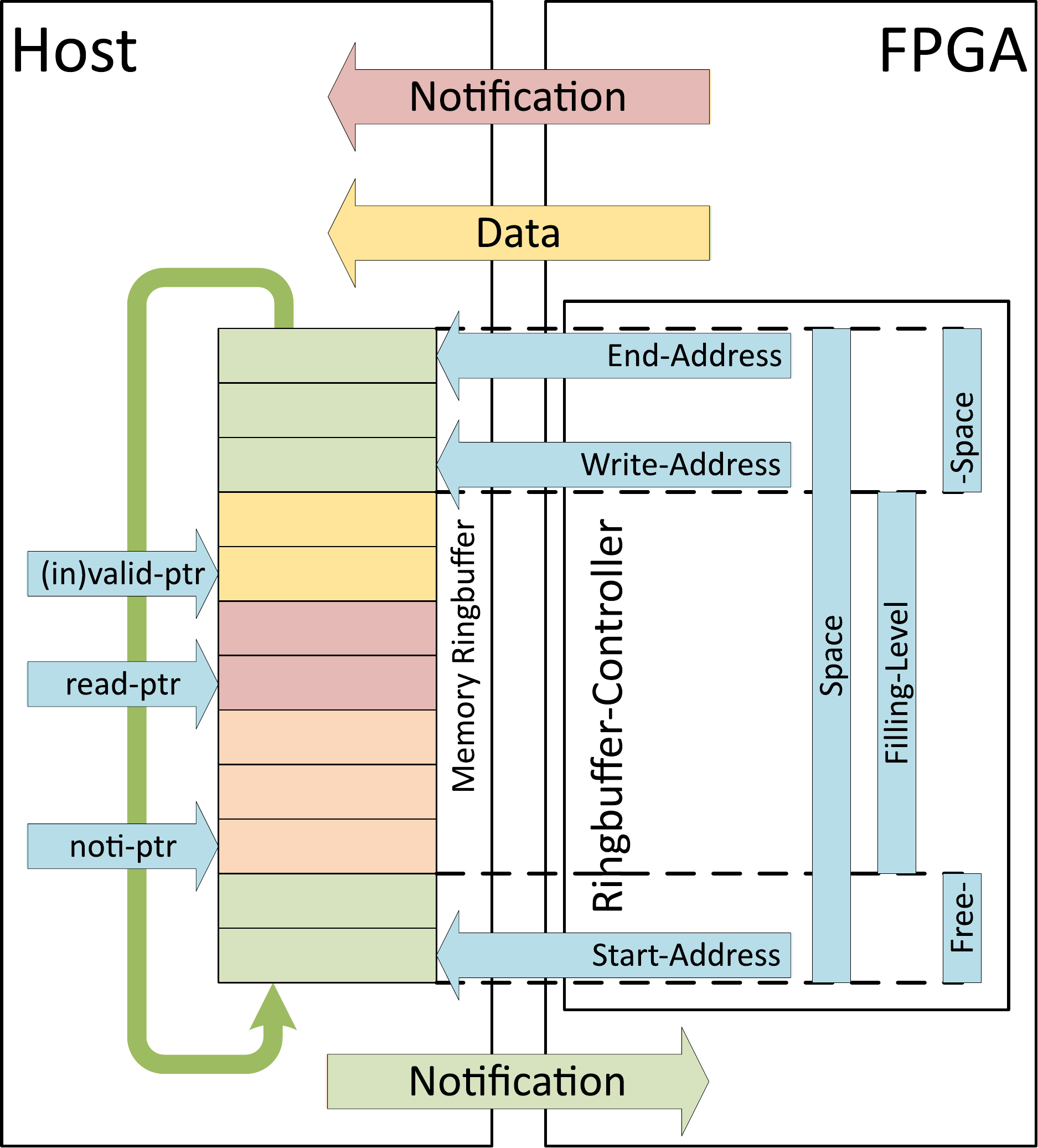}
        \label{subfig:ringbuffer}
    }
    \subfloat[]{
        \includegraphics[width=0.3\textwidth]{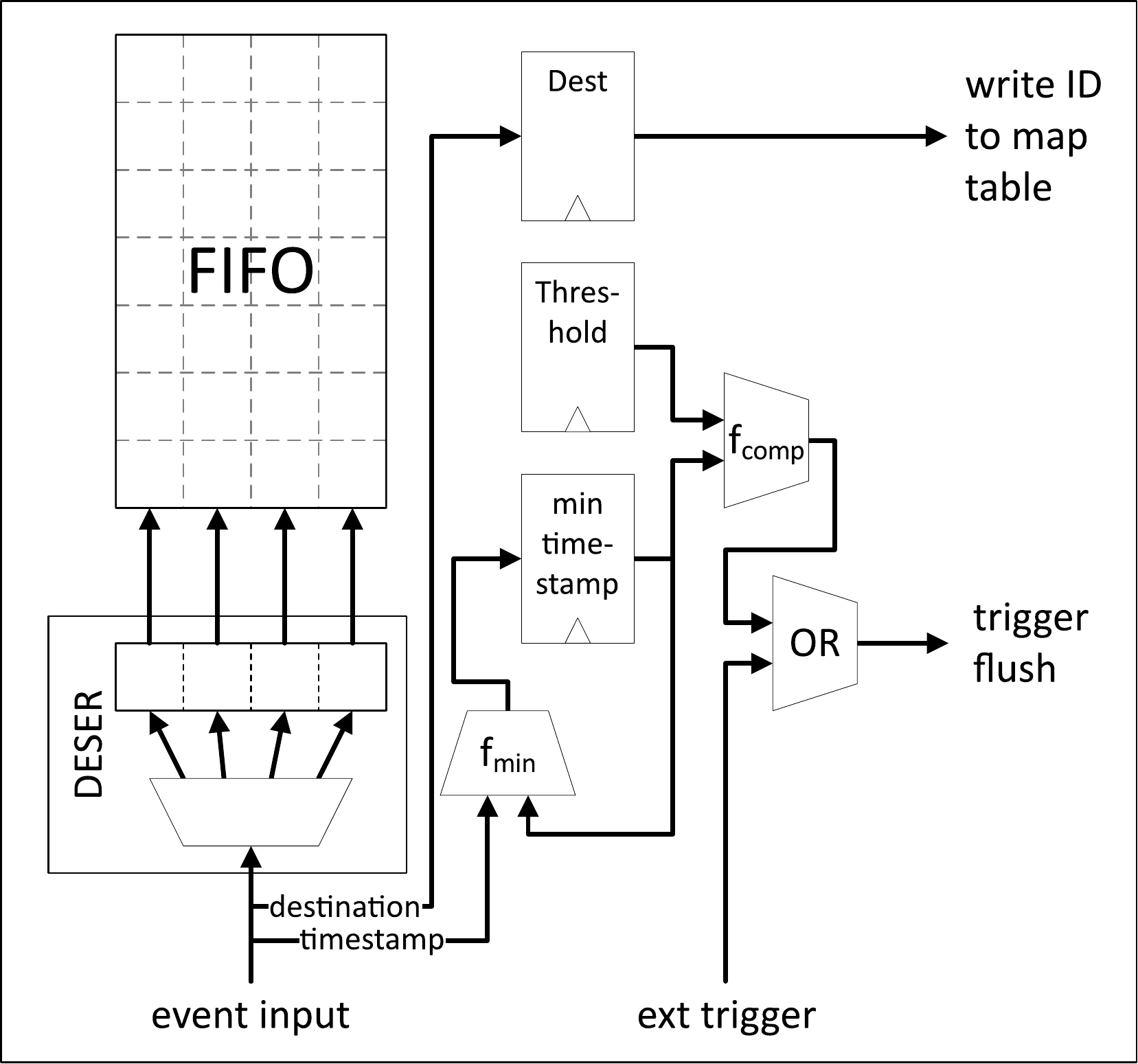}
        \label{subfig:bucket_general}
    }
    \subfloat[]{
        \includegraphics[width=0.35\textwidth]{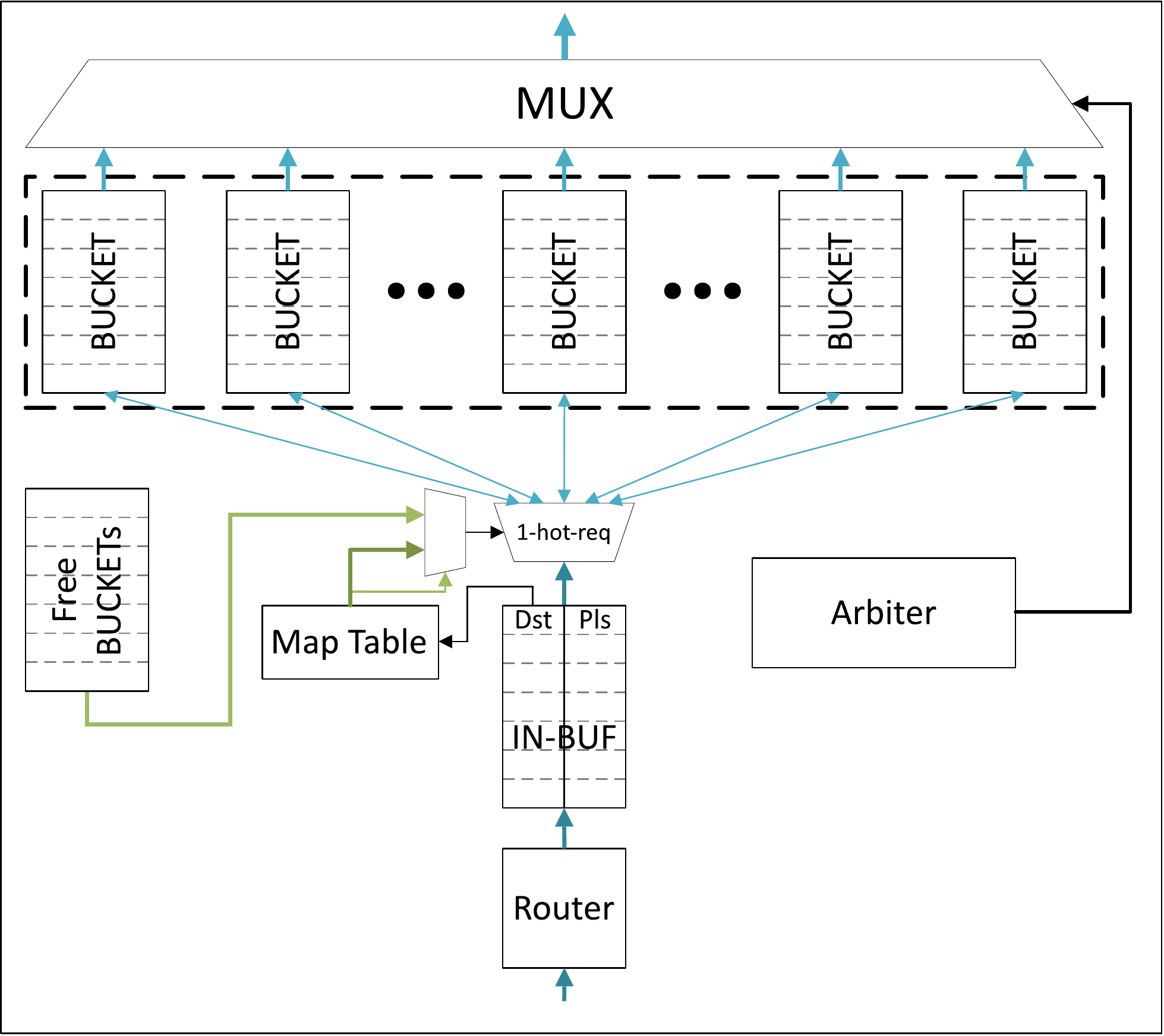}
        \label{subfig:bucket_managment}
    }
    \caption{
        \subref{subfig:ringbuffer} The ring-buffer communication scheme implemented in the FPGA for data movement to the host.
        \subref{subfig:bucket_general} The  event-accumulation buffer, called bucket. Events are deserialised to groups of four to be transmitted to the network.
        \subref{subfig:bucket_managment} Management of multiple buckets using a map table and free bucket list. The Arbiter selects the most urgent bucket for flushing.
    }
    \label{fig:ringbuffer}
\end{figure*}

\section{Communication FPGA -- Host}

The communication between FPGAs and a host node uses the Remote Memory Access (RMA) protocol \cite{nussle2009resource}.
Data moving towards an FPGA is immediately consumed by functional units in the FPGA.
Data moving back to the host is written to main memory in the host.
The arrival of new data at the host is notified to the software by making use of the
notification system in the Extoll RMA unit and the low-level driver software.

\subsection{Ring-Buffer Communication}

In order to avoid additional handshake messages,
FPGAs write their data to host memory in a predefined ring-buffer range for software processing.
This communication is conducted after the scheme shown in Figure~\ref{subfig:ringbuffer}.
The ring-buffer is always tracked by FPGA logic through the use of a write pointer and space registers.
FPGAs exchange notifications with the software,
informing each other about the amount of data written to or processed from memory.
This implements a kind of credit based flow control \cite{barkey1998credit}.

\section{Communication FPGA -- FPGA}

When wafer modules are interconnected to larger neural networks,
spike events have to be exchanged between FPGAs in the network.
Events coming from the HICANN chips to the FPGA comprise a \SI{12}{\bit} source neuron pulse address and a \SI{15}{\bit} timestamp,
stating an arrival-deadline in system-time units \cite{scholze2012dnc}.
As this does not inherently define a destination in the overall network,
a lookup table is indexed to retrieve the respective network destination-address and a generic
Global Unique Identifier (GUID) that will be transmitted over the network together with the event itself.
At the destination, another lookup table is indexed with the received GUID,
yielding a multicast mask to distribute the event among the HICANN chips connected to that FPGA.

\subsection{Event Aggregation}

Events arrive at the FPGA from the \num{8} HICANN chips with rates of up to approximately one event per \SI{210}{\mega\hertz} FPGA clock.
Due to header-overhead, single \SI{30}{\bit} events, i.e. one event per message,
can only be shifted out at a rate of one event every two clocks.
This shortcoming can be abated using packet aggregation \cite{rajkumar2008packet}.
In particular, this means accumulating events for the same destination and thereby aggregating bigger messages for transmission.
At maximum, an Extoll packet can transport up to \SI{496}{\byte} of payload, corresponding to \num{124} events.
Figure~\ref{subfig:bucket_general} shows a bucket buffer which can aggregate events until a flushing condition is met.
This is the case when the most urgent timestamp deadline is exceeded or when the buffer is full.
In addition a flush can also be triggered by external logic, managing a set of buckets.
To avoid large latencies, concurrent flushing and aggregation is implemented. Two counters track the filling level of a bucket.
One increments for incoming events while the other one decrements for flushed events.
The counters are swapped when a flush is triggered.
As there are up to $\num{2}^{\num{16}}$ possible network destinations,
the accumulation buffers need to implement a bucket renaming principle,
in analogy to the well-known register renaming \cite{cocke1991instruction}.
To always select the right buffer for an event with given destination,
the buckets are managed by a map table and a list of free buckets.
When the lookup table indicates an address to be new to the set of buckets,
the address is assigned to the next free bucket. If no bucket is free the next appropriate one is flushed.

\section{Summary and Outlook}

We have presented a low latency, high bandwidth communication strategy for a waferscale neuromorphic computing system.
Wafer modules can be interconnected by communicating spike events between FPGAs connected to HICANN chips.
This is done using the Extoll networking technology. So far, the implementation of FPGA to FPGA communication is still work in progress,
while the communication between host and FPGAs is currently being tested in the lab.
The next step is to develop a simulation model of the event aggregation buckets and verify their functionality.
One of the first multi-wafer networks will be a full scale cortical microcircuit model \cite{PotjansDiesmann2012, albada2018column}.

\section*{Acknowledgments}

We thank all present and former members of the Electronic Vision(s) and Computer Architecture research groups
contributing to the BrainScaleS Extoll communication hardware and software as well as the Extoll company for their technical support with their hardware.
This work has received funding from the European Union Seventh Framework Programme ([FP7/2007-2013]) under
grant agreement no 604102 (HBP), 269921 (BrainScaleS), 243914 (Brain-i-Nets), the Horizon2020 Framework Programme ([H2020/2014-2020])
under grant agreements 720270 (HBP) and 785907 (HBP SGA2).
This work is also funded by the Deutsche Forschungsgemeinschaft (DFG, German Research Foundation) under Germany's Excellence Strategy EXC 2181/1 - 390900948 (the Heidelberg STRUCTURES Excellence Cluster).

\bibliographystyle{IEEEtran}
\bibliography{references}

\end{multicols}

\end{document}